\documentclass[11pt,a4paper]{article}
\usepackage{epsfig}

\input epsf
\epsfverbosetrue
\newcommand{\xgo}{\mbox{$x_{\gamma}^{\rm obs}$}}

\begin{document}
\title {\begin{flushleft}{\large \tt DESY 00-166\\ November 2000}\end{flushleft}
\vspace{1cm}
\rm \bf\LARGE  Measurement of open beauty production in photoproduction at HERA \\
\vspace{1cm}}
                    
\author{ZEUS Collaboration}

\date{}

\maketitle
\begin{abstract}
\noindent
The production and semi-leptonic decay of heavy quarks have been studied in the 
photoproduction process $e^+p~\rightarrow~e^+~+~\mbox{dijet}~+~e^-~+~X$ with the ZEUS 
detector at HERA using an integrated luminosity of 38.5~${\rm pb^{-1}}$. Events with 
photon-proton centre-of-mass energies, $W_{\gamma p}$, between 134 and 269 GeV and a 
photon virtuality, $Q^2$, less than 1 ${\rm GeV^2}$ were selected requiring at least two 
jets of transverse energy $E_T^{\rm jet1(2)}~>7(6)$~GeV and an electron in the final 
state. The electrons were identified by employing the ionisation energy loss measurement. 
The contribution of beauty quarks was determined using the transverse momentum of the 
electron relative to the axis of the closest jet, $p_T^{\rm rel}$. The data, after 
background subtraction, were fit 
with a Monte Carlo simulation including beauty and charm decays. The measured beauty 
cross section was extrapolated to the parton level  with the $b$ quark restricted to the 
region of transverse momentum $p_T^{b}~>~p_T^{\rm min}~=$~5~GeV and pseudorapidity 
$|\eta^{b}|~<$~2. The extrapolated cross section is 
$1.6~\pm~0.4~(stat.)^{+0.3}_{-0.5} (syst.) ^{+0.2}_{-0.4} (ext.)~\mbox{nb}$. The result 
is compared to a perturbative QCD calculation performed to next-to-leading order.
\end{abstract}

\pagestyle{plain}
\thispagestyle{empty}
\clearpage

\pagenumbering{Roman}                                                           
\def\3{\ss}                                                                                        
                                                   %
\begin{center}                                                                                     
{                      \Large  The ZEUS Collaboration              }                               
\end{center}                                                                                       
  J.~Breitweg,                                                                                     
  S.~Chekanov,                                                                                     
  M.~Derrick,                                                                                      
  D.~Krakauer,                                                                                     
  S.~Magill,                                                                                       
  B.~Musgrave,                                                                                     
  A.~Pellegrino,                                                                                   
  J.~Repond,                                                                                       
  R.~Stanek,                                                                                       
  R.~Yoshida\\                                                                                     
 {\it Argonne National Laboratory, Argonne, IL, USA}~$^{p}$
% ...
\newpage                                                                                           
% ...
                                                           %
                                                           %
\newpage   
                                                           %
                                                           %
\begin{tabular}[h]{rp{14cm}}                                                                       
$^{a}$ &  supported by the Natural Sciences and Engineering Research                               
          Council of Canada (NSERC)  \\                                                            
$^{b}$ &  supported by the FCAR of Qu\'ebec, Canada  \\                                            
$^{c}$ &  supported by the German Federal Ministry for Education and                               
          Science, Research and Technology (BMBF), under contract                                  
          numbers 057BN19P, 057FR19P, 057HH19P, 057HH29P, 057SI75I \\                              
$^{d}$ &  supported by the MINERVA Gesellschaft f\"ur Forschung GmbH, the                          
          Israel Science Foundation, the U.S.-Israel Binational Science                            
          Foundation, the Israel Ministry of Science and the Benozyio Center                       
          for High Energy Physics\\                                                                
$^{e}$ &  supported by the German-Israeli Foundation, the Israel Science                           
          Foundation, the U.S.-Israel Binational Science Foundation, and by                        
          the Israel Ministry of Science \\                                                        
$^{f}$ &  supported by the Italian National Institute for Nuclear Physics                          
          (INFN) \\                                                                                
$^{g}$ &  supported by the Japanese Ministry of Education, Science and                             
          Culture (the Monbusho) and its grants for Scientific Research \\                         
$^{h}$ &  supported by the Korean Ministry of Education and Korea Science                          
          and Engineering Foundation  \\                                                           
$^{i}$ &  supported by the Netherlands Foundation for Research on                                  
          Matter (FOM) \\                                                                          
$^{j}$ &  supported by the Polish State Committee for Scientific Research,                         
          grant No. 112/E-356/SPUB/DESY/P03/DZ 3/99, 620/E-77/SPUB/DESY/P-03/                      
          DZ 1/99, 2P03B03216, 2P03B04616, 2P03B03517, and by the German                           
          Federal Ministry of Education and Science, Research and Technology (BMBF)\\              
$^{k}$ &  supported by the Polish State Committee for Scientific                                   
          Research (grant No. 2P03B08614 and 2P03B06116) \\                                        
$^{l}$ &  partially supported by the German Federal Ministry for                                   
          Education and Science, Research and Technology (BMBF)  \\                                
$^{m}$ &  supported by the Fund for Fundamental Research of Russian Ministry                       
          for Science and Edu\-cation and by the German Federal Ministry for                       
          Education and Science, Research and Technology (BMBF) \\                                 
$^{n}$ &  supported by the Spanish Ministry of Education                                           
          and Science through funds provided by CICYT \\                                           
$^{o}$ &  supported by the Particle Physics and                                                    
          Astronomy Research Council \\                                                            
$^{p}$ &  supported by the US Department of Energy \\                                              
$^{q}$ &  supported by the US National Science Foundation                                          
\end{tabular}                                                                                      
                                                           %
                                                           %

\newpage
\setcounter{page}{1}
\pagenumbering{arabic}
%


\section{Introduction}

High-energy collisions between a quasi-real photon, emitted by an incoming positron, 
and a proton can lead to the production of heavy quarks. Such processes allow a test of 
perturbative QCD (pQCD) since the mass of the heavy quark provides a hard scale. 

Measurements of charm production in $\gamma p$ collisions at HERA have been 
made~\cite{dstar,h1charm} by reconstructing $D^{* \pm}(2010)$ mesons. The cross 
sections generally lie above the pQCD predictions. The study of beauty production is 
important since the heavier $b$-quark mass provides a harder scale, thus making pQCD 
calculations more reliable. However, the higher mass and the smaller electric charge 
of the $b$-quark lead to cross sections in $ep$ collisions that are typically two orders 
of magnitude smaller than that of charm. The first measurement of the cross section for 
beauty photoproduction has been performed using events with a muon and two jets~\cite{h1beauty} 
and is higher than pQCD expectations. In $p\bar{p}$ 
interactions the measured beauty cross sections~\cite{ua1,cdf,d0} significantly exceed 
the predictions.

This paper presents a measurement of beauty production in photon-proton collisions using a sample of events each containing two jets and a candidate for the electron from a 
semi-leptonic decay of a heavy quark:

\vspace{-0.5cm}
\begin{eqnarray}
e^+p \rightarrow e^+(\gamma)~p \rightarrow e^+ + \mbox{dijet} + e^- + X.
\label{reaction}
\end{eqnarray}

Reaction~(\ref{reaction}) is isolated by statistically subtracting the hadronic background 
from an electron-enriched sample of events selected using measurements of the ionisation 
energy loss of charged particles.


\section{Experimental conditions}

The data used were collected by the ZEUS experiment during the 1996 and 
1997 running periods, when HERA operated with protons of energy 
$E_p~=~820~{\mbox{GeV}}$ and positrons of energy $E_e~=~27.5$~GeV. The data 
for this study correspond to an integrated luminosity of $38.5\pm0.6~\mbox{pb}^{-1}$. 
A detailed description of the ZEUS detector can be found elsewhere~\cite{ZEUS_1,ZEUS_2}.
A brief outline of the components that are most relevant for this analysis is given below.

Charged particles are measured in the central tracking detector (CTD)~\cite{ref:ctd}, 
which operates in a magnetic field of 1.43~T provided by a thin superconducting coil. 
The CTD consists of 72 cylindrical drift chamber layers, organised in 9 superlayers 
covering the polar angle\footnote{The ZEUS co-ordinates form a right-handed system with 
positive-$Z$ in the proton beam direction and a horizontal $X$-axis pointing towards the 
centre of HERA. The nominal interaction point is at $X=Y=Z=0$.} region 
$15^\circ  < \theta < 164^\circ $. The transverse-momentum resolution for 
full-length 
tracks is $\sigma_{p_T}/p_T~=~0.0058p_T \oplus 0.0065 \oplus 0.0014/p_T$ with $p_T$ in 
GeV. The energy loss of charged particles per unit track length, $dE/dx$, is also measured in 
the CTD~\cite{deppe}. 

The high-resolution uranium-scintillator calorimeter (CAL)~\cite{ref:cal} consists of 
three parts: the forward (FCAL), the barrel (BCAL) and the rear (RCAL) calorimeters. Each 
part is subdivided transversely into towers and longitudinally into one electromagnetic (EMC) 
and either one (in RCAL) or two (in BCAL and FCAL) hadronic sections (HAC). The smallest 
subdivision of the calorimeter is called a cell. The electromagnetic section of the BCAL 
(BEMC) consists of cells of $\sim$20~cm length azimuthally and mean width 5.5~cm in the 
$Z$ direction at a mean radius of $\sim$1.3~m from the beam line. These cells have a 
projective geometry as viewed from the interaction point. The CAL relative energy resolutions, 
as measured under test beam conditions are $18\%/\sqrt{E}$ for electrons and $35\%/\sqrt{E}$ 
for hadrons ($E$ in GeV). Energy deposits in the CAL were used to measure the transverse 
energy and direction of jets. Cell clusters were also formed, which were then used to 
aid in the identification of electrons for the cross-section measurements. 

The luminosity was measured from the rate of the bremsstrahlung process 
$e^+p~\rightarrow~e^+\gamma p$, where the photon was measured in a 
lead-scintillator 
calorimeter~\cite{ref:lumi} located at $Z=-107$~m.

A three-level trigger system was used to select events online~\cite{ZEUS_2,dijet_98}. 
At the third level, a cone algorithm was applied to the calorimeter cells and jets 
were reconstructed using the deposited energy and positions of the CAL cells. Events with 
at least two jets, each of which satisfied the requirements that the transverse energy  
exceeded 4 GeV and pseudorapidity was less than 2.5, were accepted.


\section{Analysis}

\subsection{Offline cuts and event selection}

To suppress backgrounds from beam-gas interactions, cosmic rays and from deep inelastic 
scattering, the following cuts were applied:

\begin{itemize}

\item neutral current deep inelastic scattering events with a well measured scattered 
positron candidate in the CAL were removed by cutting on the inelasticity, 
$y$,~\cite{dis_removal} which is the fraction of the electron energy carried by the 
photon in the proton rest frame. For an incoming positron of energy $E_e$, $y$ is 
estimated from $y_e~=~1~-~\frac{E_e^\prime}{2E_e}~(1~-~\cos\theta_e^\prime)$ where 
$E_e^\prime$ and $\theta_e^\prime$ are the energy and polar angle of the outgoing positron.
 
\item the requirement $0.2~<~y_{\rm JB}~<~0.8$ was imposed, where $y_{\rm JB}$ is the 
estimator of $y$ measured from the CAL energy deposits according to the Jacquet-Blondel 
method~\cite{YJB}. The cut was imposed after correction for energy loss due to inactive 
material in the detector. This range in $y_{\rm JB}$ corresponds to a photon-proton 
centre-of-mass energy, $W_{\gamma p}$ from 134 to 269 GeV. 

\end{itemize}

These cuts restrict the photon virtuality, $Q^2$, to less than 1~GeV$^2$. The median 
value of about $10^{-3}$~GeV$^2$ was estimated from a Monte Carlo (MC) simulation.

Jets were reconstructed using the KTCLUS~\cite{Mike_93} finder in its ``inclusive'' 
mode~\cite{Ellis_93}. KTCLUS is a clustering algorithm which combines objects with 
small relative transverse energy into jets. The objects input to the jet algorithm 
may be hadrons in a simulated hadronic final state, the final-state partons of a pQCD 
calculation, or energy deposits in the CAL. After applying the KTCLUS jet algorithm to 
the calorimeter cells, the jet transverse energy was corrected for energy loss due to 
inactive material in the detector as a function of 
$\eta^{\rm jet}_{\rm CAL}$ and $E_{T, \rm CAL}^{\rm jet}$ as described in a previous 
ZEUS publication~\cite{dijet_98}. After these corrections, all jets with 
$|\eta^{\rm jet}|~<~2.4$ and a transverse energy $E_T^{\rm jet}~>~6~$~GeV were kept. 
Each event was required to have at least two jets satisfying these criteria, with at 
least one jet having $E_T^{\rm jet}$ of more than 7 GeV.

\subsection{Electron finding}

Electrons were identified in the CTD from the $dE/dx$ of charged tracks, using a method of statistical subtraction~\cite{nikhef,wing}. Two samples of tracks were defined using 
information from the CAL: the first enriched with electrons with a background of hadrons 
(electron-enriched sample) and the second containing only hadrons (hadronic sample). The 
two samples were obtained by considering clusters in the calorimeter and performing a 
selection on the basis of the energies deposited in the EMC and HAC sections of the CAL. 
The CAL clusters matched to tracks were required to have a distance of closest approach, 
$d$, to the track of less than 20~{\rm cm}. 

\subsubsection{Electron-enriched and hadronic samples}
\label{sec:cal}

Using the matched track-cluster pairs, the electron-enriched and hadronic samples were 
defined by a cut on the ratio of electromagnetic energy, $E_{\rm EMC}$, to total energy, 
$E_{\rm TOT}$, of the clusters. The electron-enriched sample\footnote{Since the value of 
$dE/dx$ is particle dependent, the relative fractions of pions, kaons and protons determine 
the background in the electron- or positron-enriched sample. This background can only be
estimated using the hadronic sample. Monte Carlo studies showed that, for positively 
charged particles, the $\pi/(K,p)$ ratio of the hadronic sample was markedly different to 
that of the hadronic background in the positron-enriched sample. This effect is caused by the
differing cross sections for positive, compared to negative, low-energy pions, kaons and protons 
interacting with
nuclei~\cite{PDG} Therefore, since a reliable statistical subtraction was not possible, 
positrons were not considered further in this analysis.}
was required to have $E_{\rm EMC}/E_{\rm TOT}~>~0.9$, while the hadronic sample was required 
to have $E_{\rm EMC}/E_{\rm TOT}~<~0.4$, with a further requirement on the energy deposited 
in the hadronic section of the calorimeter, $E_{\rm HAC}~>~0.3$~GeV. These requirements on 
the hadronic sample efficiently rejected electrons; the residual contamination from electrons 
was estimated from photons converting into electron-positron pairs, reconstructed as 
described below, to be less than $0.03\%$. The selection criteria for the electron-enriched 
sample were $75\%$ efficient for tagging electrons, as determined from the same sample of 
photon conversions. 

\subsubsection{Measurement of \boldmath $dE/dx$}
\label{sec:dedx}

Charged particles traversing the CTD lose energy primarily by ionising gas in the detector. 
In order to estimate $dE/dx$ for each track, the truncated mean of the anode-wire pulse 
heights was calculated, which removes the lowest $10\%$ and at least the highest $30\%$ 
depending on the number of saturated hits. For electrons traversing all superlayers, 
implying a maximum of 72 
possible hits,on average 32 pulse-height measurements were retained.

Since the CTD operates at ambient atmospheric pressure, the $dE/dx$ calibration was changing 
throughout the measurement period. The measured $dE/dx$ values were corrected~\cite{wing} 
by normalising to the average $dE/dx$ for tracks around the region of minimum ionisation 
for pions, $0.3~<~p^{\rm trk}~<~0.4$~GeV, where the separation from other 
types of particles is particularly good. Henceforth $dE/dx$ is quoted in units of 
\emph{mips} - minimum ionising particles.

The measured $dE/dx$ value also depends on the polar angle. There is a trivial 
$1/\cos\theta$ dependence due to the path length which is corrected for in the subsequent 
plots. In addition, there is a dependence on $\theta$ arising from the reduction in gain 
that occurs through screening of ions in the avalanche. This effect was studied using a 
sample of electrons originating from photons which converted in the beam-pipe via the 
$\gamma~\rightarrow~e^+e^-$ process. The candidate tracks were initially selected on the basis 
of their distance of closest approach, vertex position and invariant mass. The quality 
factor,  $D = \sqrt{(\Delta xy / \sigma_{xy})^2 + (\Delta \theta / \sigma_\theta)^2}$, was 
calculated, where $\Delta xy$ is the separation of the tracks in two dimensions at the point at which their 
tangents are parallel, $\Delta\theta$ is the difference in polar angles and $\sigma_{xy}$ 
and $\sigma_\theta$ are the respective resolutions~\cite{nikhef}. The distribution of the 
quantity $D$ is shown in Fig.~\ref{fig:convert}a, which demonstrates that real conversions, 
with net charge zero, tend to have lower values of $D$ and as shown in 
Fig.~\ref{fig:convert}b, low invariant masses of the electron-positron 
pair, $M_{e^+e^-}$ . To achieve a pure sample of 
electrons, relatively hard cuts of $D~<~5$ and $M_{e^+e^-}~<~0.025$~GeV were applied. Electron-positron candidates of net-zero and 
net-two units of charge were considered and the two subtracted as shown in 
Fig.~\ref{fig:spq}a. The $dE/dx$ distribution for the sample of clean electrons has a 
roughly Gaussian shape centred about $dE/dx~\sim~1.4$~mips with width 0.14~mips, 
corresponding to a resolution of $\sim10\%$. 
Figure~\ref{fig:spq}b shows that the $dE/dx$ value exhibits a strong dependence on the polar 
angle, $\theta^{\rm trk}$, as expected from the space-charge effect~\cite{spq}. The value 
of $dE/dx$ at $\theta^{\rm trk}~=~\pi/2$ is about $10\%$ lower than the most forward and 
backward values in the range of $\theta^{\rm trk}$ considered. Using this sample of electrons, correction factors 
were obtained which depend on $\theta^{\rm trk}$. The $dE/dx$ for electrons was corrected 
such that the mean was 1.4 mips.

\subsubsection{Tracking requirements}

Negatively charged tracks with a transverse momentum relative to the $ep$ beam axis, 
$p_T^{\rm trk}$, greater than 1.6~GeV were selected. Electron candidates were restricted to the central 
region of the detector, $|\eta^{\rm trk}|~<~1.1$, corresponding to 
$0.64<\theta^{\rm trk}<2.50$ radians, where the resolution in $dE/dx$ is constant to 
within $10\%$. A small slice in $\eta^{\rm trk}$ was removed from the analysis, corresponding 
to the region in which tracks were matched to clusters where the BCAL and RCAL meet, where
CAL clusters were not well reconstructed. Figure~\ref{fig:dedx-p} shows that for 
$1.6~<~p_T^{\rm trk}<10$~GeV, all hadrons have an average $dE/dx$ value well below that of 
electrons, thus allowing the separation of electrons from hadrons.

\subsubsection{Background electrons from converting photons}

The major source of background to electrons from semi-leptonic decays of heavy quarks (prompt 
electrons) comes from photon conversions in the detector. Electron-positron pairs were 
initially selected as discussed in Section~\ref{sec:dedx}. A loose cut of $D <15$ was chosen.

Tracks above a momentum of 0.2 GeV were reconstructed with an efficiency of greater 
than $95\%$. The number of potential conversion candidates not identified after 
this cut because the electron-positron pair was asymmetric in momentum was determined from a 
calculation of pair production~\cite{tsai} rather than by relying on a MC model. After the 
$p^{\rm trk}~>~0.2~{\rm GeV}$ cut on the positron candidate was imposed, good agreement was 
seen between data and expectation for the shape of $E_{e^-}/E_\gamma$, the fraction of the photon's energy 
carried by the electron, as shown in Fig.~\ref{fig:tsai}. This 
demonstrates that the identified electron-positron candidates originated from photon 
conversions. The requirement for the positron's momentum, $p^{e^+}>0.2~{\rm GeV}$, was 
then removed in the calculation and the ratio, 
$\epsilon(E_\gamma)=N(p_T^{e^-}>1.6~{\rm GeV},~p^{e^+}>0.2~{\rm GeV}) 
                           / N(p_T^{e^-}>1.6~{\rm GeV})$
of the two theoretical predictions (the solid and dashed curves in Fig.~\ref{fig:tsai}) 
determined, where $N$ is the number of photon conversions. The efficiency, $\epsilon$, 
averaged over the photon energy was found to be 83$\%$.

The overall efficiency for tagging the background due to electrons from photon conversions 
was determined from MC simulations. This was achieved by performing the analysis procedure 
with an inclusive MC conversion sample but demanding no prompt electron. The number of 
electrons found by the analysis procedure, together with the number of electrons identified 
as coming from converting photons by the above two steps, was determined. The ratio of the 
number of identified conversion candidates to the total number of candidates passing the 
analysis cuts was $90\pm3\%$ which in combination with $\epsilon=83\%$, leads to an overall 
efficiency for the identification of conversion candidates of $75\pm3\%$.

\subsubsection{Electrons from Dalitz decays, \boldmath $\pi^0 \rightarrow e^+e^-\gamma$}
\label{sec:dalitz}

A substantial background is produced by Dalitz decays, $\pi^0 \rightarrow e^+e^-\gamma$. 
To remove this background, all tracks in the electron-enriched sample that were not 
identified as coming from conversions, were combined with tracks of positive charge and 
the invariant mass formed. The positively charged tracks were required to have originated 
from the primary vertex, to have passed through at least the first three superlayers of 
the CTD and to have $p_T^{\rm trk}~>~0.1$~GeV. Since MC simulations showed that no fully 
reconstructed Dalitz decays survived a cut of $M_{e^+e^-}>0.2$~GeV, an electron candidate 
was removed if any combination failed this cut. The overall rejection efficiency was 
$84\pm2\%$.

\subsubsection{Electron signal}
\label{sec:signal}

In order to subtract statistically the hadronic background from the electron-enriched sample, 
the dependence of the measured $dE/dx$ on momentum and polar angle must be taken into 
account. As the $p$ and $\theta$ distributions for the electron-enriched and hadronic samples 
differ, subtraction of the hadronic background was carried out by reweighting the hadronic sample to 
give the same distribution as the hadronic background in the electron-enriched sample. This 
was achieved by binning the distributions in transverse momentum and pseudorapidity and 
normalising the hadronic sample to the electron-enriched sample in the region 
$0.5<dE/dx<1.1$~mips. In this analysis, 64 bins, with eight divisions in $p_T^{\rm trk}$ and 
eight in $\eta^{\rm trk}$, were used.

Figure \ref{fig:signal}a shows the $dE/dx$ distributions for both the electron-enriched and 
hadronic samples. The subtracted distribution together with the background from the photon 
conversions are shown in Fig.~\ref{fig:signal}b. Similar distributions are seen in 
the individual bins of $p_T^{\rm trk}$ and $\eta^{\rm trk}$. In Fig.~\ref{fig:signal}a, the two 
distributions have the same shape for $dE/dx$ values below $\sim~1.1$ mips, but there is a 
clear excess of the electron-enriched sample 
over the hadronic sample at larger values of $dE/dx$, indicating the presence of 
electrons. The excess over the conversion signal was used to extract cross sections 
for electrons from heavy-quark decays. A cut at the mean value for electrons of 
$dE/dx = 1.4$ mips was made, which, assuming a symmetrical Gaussian distribution (as suggested by
Fig.~\ref{fig:spq}a), has an efficiency of $50\%$. Varying the cut in steps of 0.01 mips between 
1.3 and 1.5 mips gave the same results within the statistical uncertainties. Cutting at lower 
values of $dE/dx$ gave no overall improvement in the statistical error.

The number of identified electrons after this procedure is 1480~$\pm$~63, of which 
537~$\pm$~29 were attributed to photons converting into electron-positron pairs, resulting 
in  943~$\pm$~69 electrons used for the cross section determination. The same 
statistical subtraction procedure was used in each bin to extract the differential cross 
sections.
 

\section{Monte Carlo event simulation}
\label{sec:mc}

The acceptance and the effects of detector response were determined using samples of MC 
events. The programs {\sc Herwig~5.9}~\cite{herwig} and {\sc Pythia~5.7}~\cite{pythia}, 
which implement the leading-order matrix elements followed by parton showers, were used. 
For all generated events, the ZEUS detector response was simulated in detail using a 
program based on GEANT3.13~\cite{geant}.

At leading order (LO), two types of processes can be distinguished: direct photon processes, 
in which the photon couples directly to a parton in the proton; and resolved photon 
processes, where the photon acts as a source of partons, one of which participates in the 
hard interaction. Samples of direct and resolved photon events, including heavy-quark 
excitation processes, were generated separately. For acceptance corrections, the MC events 
were generated with the CTEQ-4D~\cite{cteq} structure function for the proton and 
GRV-LO~\cite{grv} for the photon. The default quark masses were used in both {\sc Herwig} 
and {\sc Pythia}.

For the fitting procedure and MC predictions, the GRV94-LO structure function for the proton 
and the GRV-LO for the photon were used. The quark masses were set to the nominal values of 
$m_b~=~4.75$~GeV and $m_c~=~1.5$~GeV. Samples with different input parton density functions 
and different quark masses were used to evaluate systematic effects.

Both the shape and normalisation of the MC samples were compared to the differential cross 
sections. Fits to the data yielded the fraction of resolved photon processes as well as of 
beauty production. The cross-section value predicted by the MC model was also compared to the 
measured $b~\rightarrow~e^-$ cross section. The MC predictions were then used to extrapolate the measured 
cross section to the parton level to facilitate a comparison with an NLO calculation.


\section{Event characteristics}

Comparisons of the distributions of kinematic quantities between the data and the {\sc Herwig} 
and {\sc Pythia} MC simulations, which include production of all five flavours of quark, 
are shown in Fig.~\ref{fig:datamc}. Both MC models describe the data reasonably well. 

The number of jets is more accurately described by the {\sc Pythia} MC but the jet quantities 
themselves are well described by both MC programs. The general shape of the $p_T^{\rm trk}$ distribution 
is well described, although there are slightly more events in the data at 
high $p_T^{\rm trk}$. The matching of the track-cluster pairs is also well described, as 
demonstrated by the distribution of $d$, the distance of closest approach of the
track and the cluster. The separation of the jet and electron candidate 
in $\eta-\phi$ space, $R_{\rm e-jet}$, is peaked at low values with a flat tail at 
$R_{\rm e-jet}~>~1$, amounting to $10\%$ of the sample; it is described reasonably by the MC programs. 
Confidence in the use of the MC programs for acceptance corrections was thus justified.

Also shown in Fig.~\ref{fig:datamc} is the contribution in the MC model from the semi-leptonic 
decays of beauty quarks. These events, in general, have jets with larger transverse energy 
and have a tendency to be more forward (i.e. nearer to the proton beam direction) in 
$\eta^{\rm jet}$. The prompt electron is more 
separated from the jet, being produced with a higher transverse momentum and also in a 
more forward direction.

\section{Cross section measurements\label{sect:results}}

\begin{figure}[htb]
\centerline{\epsfig{file=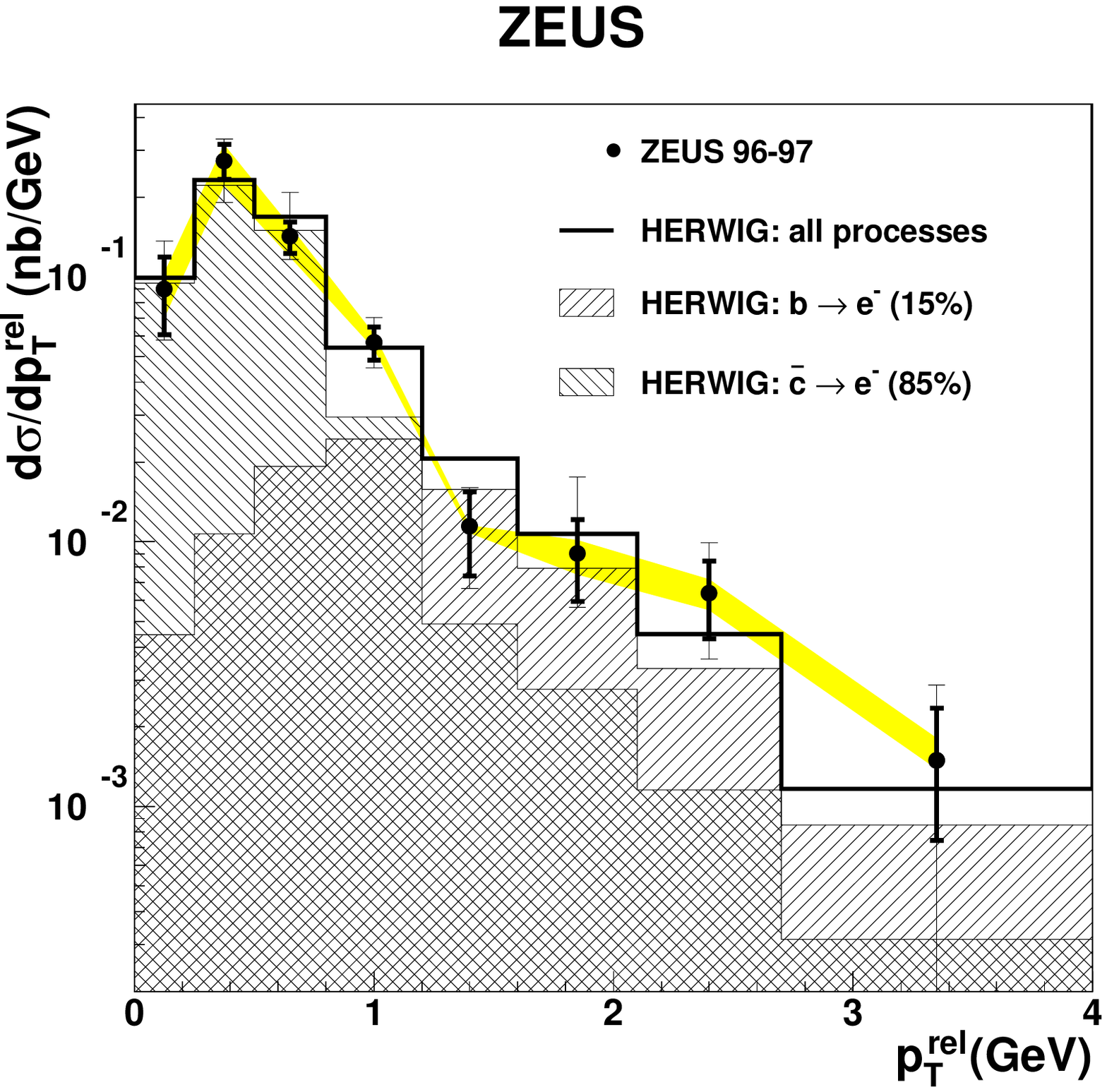}}
\caption{The differential cross-section $d\sigma/dp_T^{\rm rel}$ for heavy quark decays. 
The inner error bars represent statistical errors and the outer bars statistical and 
systematic errors added in quadrature. The effect of the energy scale uncertainty is shown as the 
shaded band. The experimental data is compared to the {\sc Herwig} MC 
prediction (solid line) for all quark flavours, 
which has been fitted to the data and scaled up by a factor of 3.8. The components
from the beauty processes (forward-diagonally hatched 
histogram) and from charm (backward-diagonally hatched histogram)
predicted by the {\sc Herwig} MC model are indicated separately.}
\label{fig:ptrel}
\end{figure}

The signal for beauty decays can be seen in Fig.~\ref{fig:ptrel}, in which the 
$p_T^{\rm rel}$ distribution, where $p_T^{\rm rel}$ is the momentum of the electron 
transverse to the axis of the jet to which it is closest, is compared to the MC prediction. 
The data peak at low $p_T^{\rm rel}$, consistent with predominantly semi-leptonic decays 
of charm quarks, with other contributions mostly from $\tau$ and $\eta$ decays contributing 
less than $3\%$. At high $p_T^{\rm rel}$, the data fall less steeply than the predictions for 
charm. The data are consistent with a significant contribution from $b$ decays.

\subsection{Differential cross sections}

Differential cross sections were determined as a function of $p_T^{\rm rel}$ and \xgo, the 
fraction of the photon's momentum contributing to the production of the two highest 
transverse energy ($E_T^{\rm jet}$) jets. The variable \xgo \ is defined as~\cite{xgo}:
 
\begin{equation}
\xgo = \frac{\sum_{\rm jets} E_T^{\rm jet} e^{-\eta^{\rm jet}}}{2yE_e},
\label{eqxgo}
\end{equation}

where the sum is over the two jets of highest transverse energy. Cross sections for 
reaction~(\ref{reaction}) were measured in a restricted kinematic region with 
$p_T^{e^-}~>~1.6$~GeV and $|\eta^{e^-}|~<~1.1$. Differential electron cross sections, 
$d\sigma/dp_T^{\rm rel}$ and $d\sigma/d\xgo$ , were determined in the region 
$Q^2~<~1~{\rm GeV^2}$, $0.2~<~y~<~0.8$, requiring events with at least two jets
with $E_T^{\rm jet1(2)}~>~7(6)$~GeV and $|\eta^{\rm jet}|~<~2.4$. For a given luminosity, ${\mathcal L}$, 
the cross section, $\sigma_i$, in bin $i$ is given by 
$\sigma_i~=~N_i^{\rm corr}~/~({\mathcal L}~\cdot~\Delta)$, where $N_i^{\rm corr}$ is the 
acceptance-corrected number of electrons in the bin $i$ and $\Delta$ is the bin width. The 
acceptance correction-factors were obtained from MC simulations using a bin-by-bin  
method. The reference MC model was {\sc Herwig}, with {\sc Pythia} used as a systematic check. 
At low $p_T^{\rm rel}$, the value of the correction factor was 2.3, decreasing with increasing 
values of $p_T^{\rm rel}$ to a minimum of 0.8. For the cross section as a function of \xgo, the 
correction factors were in the range 1.1 $-$ 2.7, increasing with increasing \xgo. 

\begin{figure}[htb]
\centerline{\epsfig{file=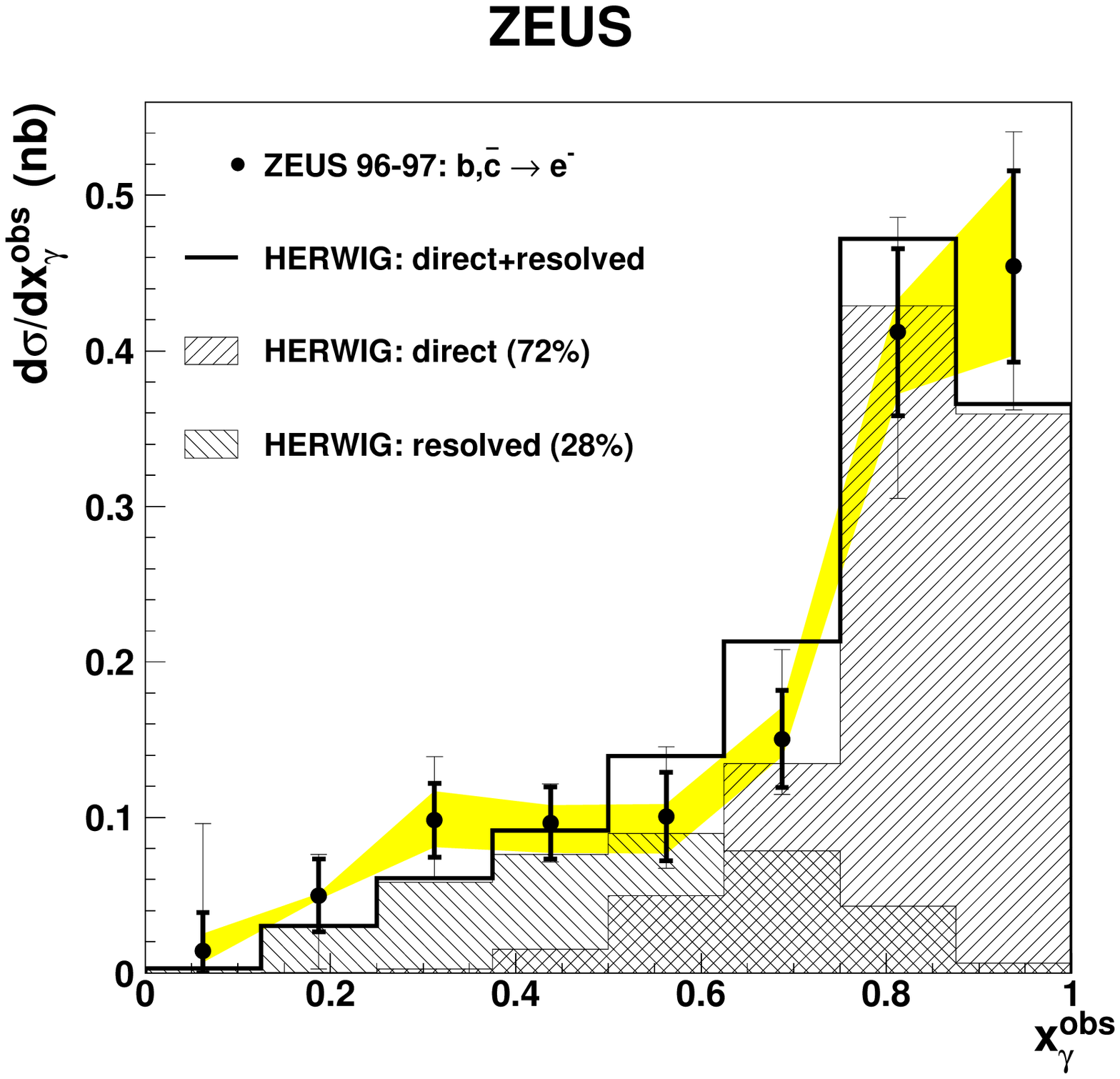}}
\caption{The differential cross-section $d\sigma/d\xgo$ for heavy quark decays. 
The inner error bars represent statistical errors and the outer bars statistical and 
systematic errors added in quadrature. The effect of the energy scale uncertainty is shown as the 
shaded band. The experimental 
data is compared to the {\sc Herwig} MC prediction (solid line) for all quark flavours, which 
has been fitted to the data 
and scaled up by a factor of 3.8. The direct (forward-diagonally hatched histogram) and resolved 
photon components (backward-diagonally hatched histogram) are indicated separately.}
\label{fig:xgamma}
\end{figure}

The measured differential cross sections, $d\sigma/dp_T^{\rm rel}$ and $d\sigma/d\xgo$, 
where the $e^-$ comes from the semi-leptonic decay of a heavy quark, are shown in 
Figs.~\ref{fig:ptrel} and~\ref{fig:xgamma} respectively. The contribution from Dalitz 
decays was removed via the procedure and with the efficiency estimated in Section~\ref{sec:dalitz}. 
The data are compared to the expectations of the {\sc Herwig} MC simulation, which was area 
normalised to the data for a comparison of shape; the scaling factor was 3.8.

Figure \ref{fig:xgamma} shows a peak at high \xgo, consistent with direct photon processes. 
However, the tail at low \xgo \ cannot be explained by direct processes alone. The {\sc Herwig} prediction of 35$\%$ resolved photon contribution (including flavour-excitation 
processes) and 65$\%$ direct gives good agreement with the data. 
Fitting the shape of the direct and resolved photon MC distributions to the data gave a 
resolved component of 28$\pm$5({\em stat.})$\%$ ($\chi^2/$ndf~=~1.5). It can therefore be 
concluded that LO MC models require resolved photon processes to describe heavy quark production, 
as was observed in the ZEUS result for charm~\cite{dstar}.
 
\subsection{Beauty cross section}

The beauty cross section was extracted by fitting the $p_T^{\rm rel}$ distribution of the 
data to the sum of contributions from beauty and charm. In the fit, the relative fraction 
of beauty and charm in the MC simulation was varied and the fraction of beauty processes in 
the kinematic region was extracted by minimising $\chi^2$. The quoted beauty cross section 
below includes only the direct semi-leptonic decay from a $b$ quark to an electron since 
the cascade decay, $\bar{b}~\rightarrow~\bar{c}~\rightarrow~e^-$ is included in the 
background expected from charm because it exhibits a $p_T^{\rm rel}$ spectrum more similar  
to a charm than a beauty decay. 

The percentage of beauty production was determined to be  $14.7~\pm~3.8~(stat.)~\%$ 
($\chi^2/$ndf~=~1.1). Using this, the cross section for beauty production 
in the restricted kinematic region: $Q^2~<~1~{\rm GeV^2}$, $0.2~<~y~<~0.8$, with at least 
two jets, $E_T^{\rm jet1(2)}~>~7(6)$~GeV and $|\eta^{\rm jet}|~<~2.4$ and a prompt electron 
with $p_T^{e^-}~>~1.6$~GeV and $|\eta^{e^-}|~<~1.1$, was found to be

\begin{equation}
\sigma^{b \rightarrow e^-} 
      _{e^+ p \rightarrow {e^+ + \rm dijet} +e^- + X} 
      = 24.9 \pm 6.4 ^{+4.2}_{-7.3} \ {\rm pb}.
\label{eq:bvis}
\end{equation}

The predictions from the two MC models are 8 pb for {\sc Herwig} and 18 pb for {\sc Pythia}, 
using the parameter settings quoted in Section~\ref{sec:mc} for the fitting procedure. The large 
difference in the MC predictions comes mainly from the different default treatments of 
$\alpha_s$, different scales and hadronisation and the use of massive matrix elements in 
{\sc Herwig}~\cite{herwig} and massless in {\sc Pythia}~\cite{pythia} for the generation 
of flavour-excitation processes. Despite the cross-section differences, the predicted 
fractions of beauty production are 16.2$\%$ in {\sc Herwig} and 19.5$\%$ in {\sc Pythia}, 
both in reasonable agreement with the data value of $14.7~\pm~3.8~(stat.)~\%$. Using 
the same procedure for charm, the measured cross section was found to be 
in reasonable agreement with the ZEUS measurement of $D^*$ production~\cite{dstar}.

\subsection{Systematic uncertainties}

A detailed study of possible sources of systematic uncertainty was performed for the 
differential and total cross-section measurements by varying cuts on the quantities to check the 
acceptance and by using the alternative MC model to check the stability of the fit. The largest 
contributions to the systematic error for the differential cross sections 
$d\sigma/dp_T^{\rm rel}$ and $d\sigma/d\xgo$ \ were due to the uncertainty in the CAL 
energy scale and the use of {\sc Pythia} instead  of {\sc Herwig} to correct the data. In the 
case of the measured beauty total cross section, the significant errors  were:

\begin{itemize}

\item using {\sc Pythia} instead of {\sc Herwig} gave an uncertainty of $-8\%$;

\item the range for $y_{JB}$ was varied by the resolution of 8$\%$, leading 
      to an uncertainty of $^{+ \ 0}_{-11}\%$;

\item requiring $R_{\rm e-jet}~<~2$ rather than no requirement gave an uncertainty of 
      $-10\%$.

\item the normalisation range of the hadronic background to the electron-enriched sample was 
      changed to $0.5~<~dE/dx~<~1.04$~mips, changing the cross section by $-12\%$.

\item varying the distance of closest approach of the track-cluster pair, $d$, by 
      $\pm~5$~cm changed the cross section by $^{+7}_{-3}\%$. 

\item varying the cuts for the hadronic sample on the fraction $E_{\rm EMC}/E_{\rm TOT}$ between 0.3 and 0.5 and for the electron-enriched sample between $E_{\rm EMC}/E_{\rm TOT}~>~0.85$ and 0.95 gave an uncertainty 
      of $^{+13}_{-16}\%$.

\item using the CTEQ-4L proton structure function yielded a systematic error of $-4\%$, while 
      using the GS96-LO~\cite{gs96} photon structure function yielded a systematic error of 
      $-5\%$. 

\end{itemize}

All systematic uncertainties on the differential cross sections, excluding the 
correlated uncertainties due to the luminosity and hadronic energy scale, were added in 
quadrature. For the total cross section, all systematic uncertainties except that due to 
the luminosity were added in quadrature. 


\section{Extrapolation to parton level cross section}

\subsection{Extrapolation procedure}

The measured beauty cross section was extrapolated to the parton level in a restricted 
range of the transverse momentum and pseudorapidity of the beauty quark using the two MC 
models. The region in pseudorapidity for the beauty quark was defined such that the 
acceptance was reasonably constant over the range considered. The MC simulation predicts 
that 95$\%$ of events used in the cross-section determination of Eq.~(\ref{eq:bvis}) have 
a $b$-quark of transverse momentum $p_T^b >$ 5 GeV. Accordingly, the measured cross section 
was extrapolated to the parton level for the region $p_T^b > p_T^{\rm min} =$ 5 GeV, 
$|\eta^{b}|~<~2$, $Q^2~<~1~{\rm GeV^2}$ and $0.2~<~y~<~0.8$. The extrapolation factor 
$f_{\rm ext}$ is then given by

\[ f_{\rm ext} = \left[ \frac{\sigma^{b \rightarrow e^-}_{e^+p \rightarrow e^+bX}} 
                             {\sigma^{b \rightarrow e^-} 
                                    _{e^+ p \rightarrow e^+ + {\rm dijet} +e^- + X}}
                                    \right]_{\rm MC}. \]

The value obtained using the {\sc Herwig} MC generator was $f_{\rm ext}~=~6.8$. To correct 
to the full $b$ cross section the branching ratio for the process of $b \rightarrow e^-$ 
$(10.73~\pm~0.35)~\%$~\cite{PDG} was used. The major source of uncertainty on the 
extrapolation factor arises from the uncertainty in the hadronisation, quantified by 
the use of {\sc Pythia}, to be -26$\%$; {\sc Pythia} uses the Lund model while {\sc Herwig} 
uses the cluster model. The other significant sources of uncertainty arise from varying 
the space-like evolution parameters\footnote{For the central analysis, the 
default setting, denoted by the variable {\tt ISPAC = 0}, allows the backward evolution 
down to some cut-off, {\tt QSPAC}, which was set to 1 GeV. By changing to {\tt ISPAC = 1}, the 
backward evolution continues to the infra-red cut-off, but the parton density functions are 
frozen at the value {\tt QSPAC}, which is again set to 1 GeV.} ($-11\%$), varying the mass 
of the $b$ quark from 4.5~GeV to 5.0~GeV ($^{+8}_{-2}\%$) and the branching ratio ($\pm 3\%$). 
Applying $f_{\rm ext}$ and the branching ratio to the measured cross section gives

\[ \sigma^{\rm ext}_{ep \rightarrow e^+bX} = 
1.6 \pm 0.4 (stat.) ^{+0.3}_{-0.5} (syst.) ^{+0.2}_{-0.4} (ext.)~\mbox{nb} \]

where the last error given is the estimate for the error arising from the extrapolation 
procedure.

\subsection{Comparison with NLO predictions}

\begin{figure}[htb]
\centerline{\epsfig{file=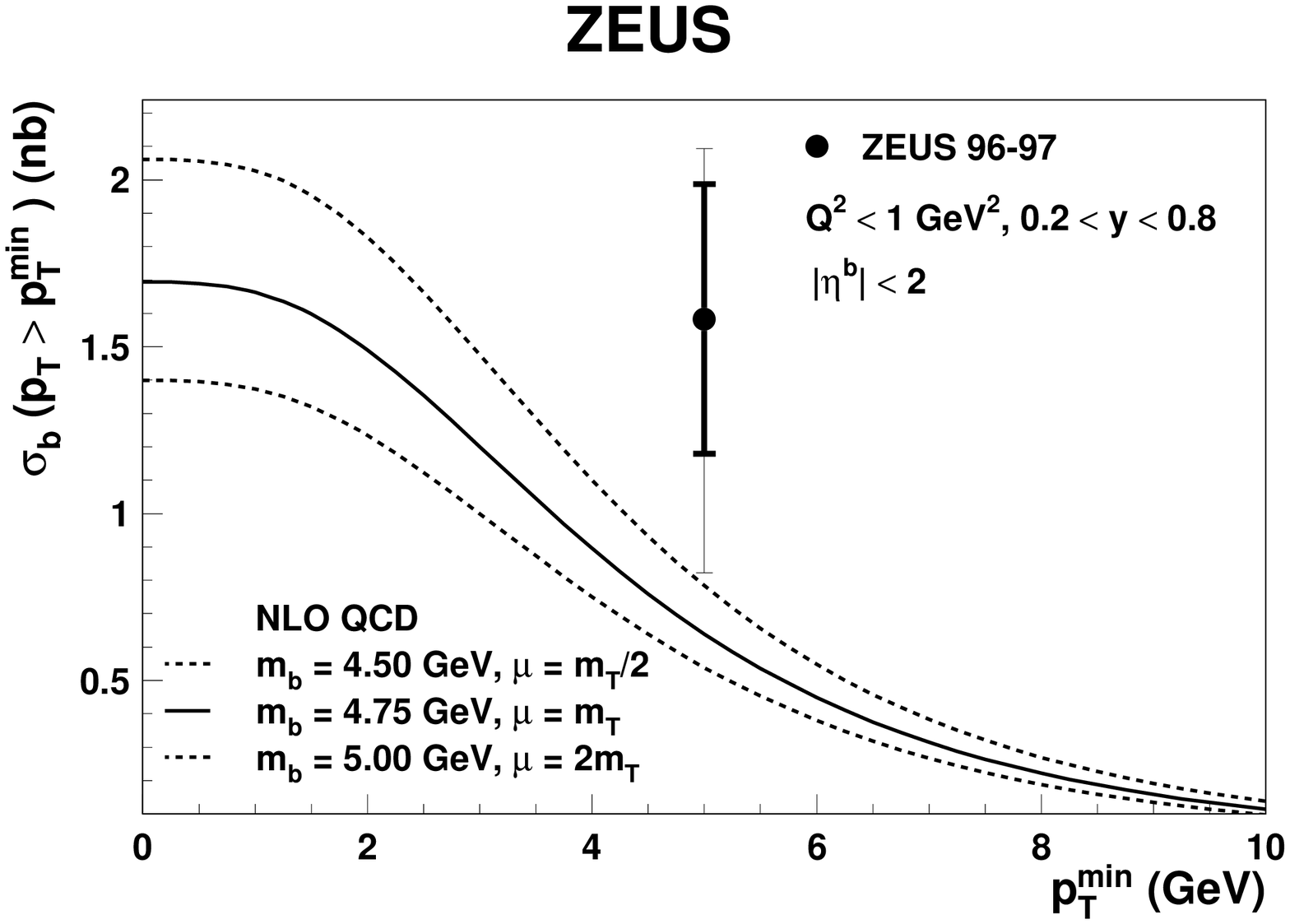,height=12cm}}
\caption{The extrapolated $b$ cross section at a fixed $p_T^{\rm min}$ value compared with 
theoretical predictions plotted as a function of $p_T^{\rm min}$. The inner error bars 
represent the statistical error and the outer bars statistical, systematic and 
extrapolation errors added in quadrature. The curves show the predictions from NLO QCD for varying $b$-quark mass and varying factorisation and
renormalisation scale $m_T = \sqrt{m_b^2 + p_T^2}$.} 
\label{fig:ptmin}
\end{figure}

The cross section is compared in Fig.~\ref{fig:ptmin} to a NLO QCD calculation~\cite{frix}. 
The prediction comes from a fixed-order calculation in which $b$ quarks are not active 
partons in the proton and photon structure functions and are generated dynamically in the 
hard sub-process. The NLO calculation uses the MRST99~\cite{mrst} and GRV-G HO~\cite{grv} 
parton-density parametrisations for the proton and photon, respectively. For the 
central prediction, the renormalisation and factorisation scales are set to the transverse 
mass, $\mu_R = \mu_F = m_T = \sqrt{m_b^2 + p_T^2}$, where $m_b~=~4.75$~GeV. The predicted 
cross section in the region $p_T^b~>~5$~GeV, $|\eta^b|~<~2$, $Q^2 < 1~{\rm GeV^2}$ and 
$0.2~<~y~<~0.8$ for the above settings is $0.64$ nb. The values 
produced by variation of the 
mass of the beauty quark (4.5 and 5.0~GeV) and the renormalisation and factorisation 
scales ($2m_T$ and $m_T/2$) are shown in Fig.~\ref{fig:ptmin} as the upper and lower 
predictions. Using other sets of parton density functions, e.g. MRST99 ($g \downarrow$), 
MRST99 ($g \uparrow$), MRST99($\alpha_s \downarrow \downarrow$) and 
MRST99($\alpha_s \uparrow \uparrow$), in which the extremes of an allowed range of values 
are taken \cite{mrst}, resulted in variations of the NLO predictions for $p_T^{\rm min}~>~5$~GeV of within 
$\pm~4\%$. Using CTEQ~5M1~\cite{cteq5} and GRV98~HO~\cite{grv98} proton structure functions changes
the prediction by $+2\%$ and $-5\%$, respectively.

The extrapolated cross section lies somewhat above the central NLO prediction, consistent with the general observation that 
NLO QCD calculations underestimate beauty production both in hadroproduction~\cite{ua1,cdf,d0} 
and photoproduction at HERA~\cite{h1beauty}.

\section{Summary}

The production and semi-leptonic decay of heavy quarks has been studied in the 
photoproduction process $e^+p~\rightarrow~e^+~+~\mbox{dijet}~+~e^-~+~X$ with the ZEUS 
detector at HERA using an integrated luminosity of 38.5~${\rm pb^{-1}}$. Events with 
photon-proton centre-of-mass energies, $W_{\gamma p}$, in the range $134~<~W_{\gamma p}~<~269$~GeV 
and a photon virtuality, $Q^2~<~1~{\rm GeV^2}$, were selected with at least two jets of 
transverse energy $E_T^{\rm jet1(2)}~>7(6)$~GeV and an electron in the final state. The 
beauty cross section was measured to be 
$\sigma^{b \rightarrow e^-}_{e^+ p \rightarrow {e^+ + \rm dijet} +e^- + X} 
= 24.9 \pm 6.4 ^{+4.2}_{-7.3} \ {\rm pb}$. This cross section was extrapolated to the 
parton level with a $b$ quark restricted to the region of transverse momentum 
$p_T^{b}~>~ 5$~GeV and pseudorapidity $|\eta^{b}|~<$~2 for the events with  
$Q^2~<~1~{\rm GeV^2}$ and $0.2~<~y~<~0.8$. The value obtained, 
$\sigma^{\rm ext}_{ep \rightarrow e^+bX}~=~1.6~\pm~0.4~(stat.)~^{+0.3}_{-0.5} (syst.)^{+0.2}_{-0.4} (ext.)~\mbox{nb}$, 
lies somewhat above the NLO predictions, in agreement with results both from hadroproduction 
and photoproduction.

\section*{Acknowledgements}
The strong support and encouragement of the DESY Directorate have been invaluable, and we 
are much indebted to the HERA machine group for their inventiveness and diligent efforts. 
The design, construction and installation of the ZEUS detector have been made possible by 
the ingenuity and dedicated efforts of many people from inside DESY and from the home 
institutes who are not listed as authors. Their contributions are acknowledged with great 
appreciation. We would like to thank S. Frixione for providing us with the program for his 
NLO calculation.

%


\newpage

\unitlength=1mm
\begin{picture}(0,0)(100,100)
\put(168,70){\bf \Large{(a)}}
\put(251,70){\bf \Large{(b)}}
\end{picture}
\begin{center}
~\epsfig{file=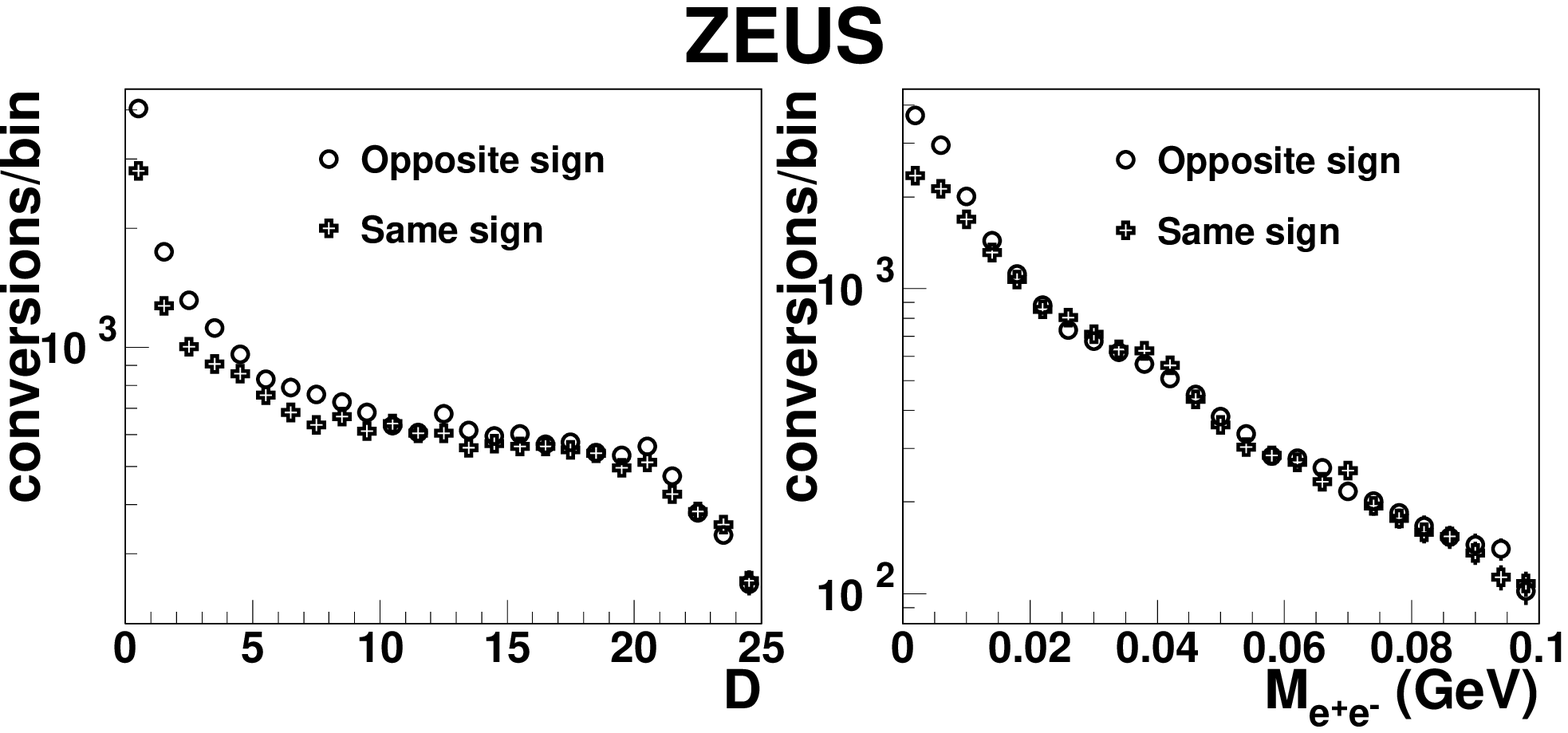,height=8.25cm}
\end{center}
\begin{figure}[htb]
\caption{(a) The distribution of the quality factor, $D$, (see Section~\ref{sec:dedx}) for 
conversion candidates. (b) the invariant mass, $M_{e^+e^-}$, for conversion candidates. 
In (a) and (b), the conversion candidates resulting in pairs 
with zero net charge are shown as 
points; those pairs having non-zero net charge are shown as the crosses.}
\label{fig:convert}
\end{figure}

\newpage

\unitlength=1mm
\begin{picture}(0,0)(100,100)
\put(165,76){\bf \Large{(a)}}
\put(240,76){\bf \Large{(b)}}
\end{picture}
\begin{center}
~\epsfig{file=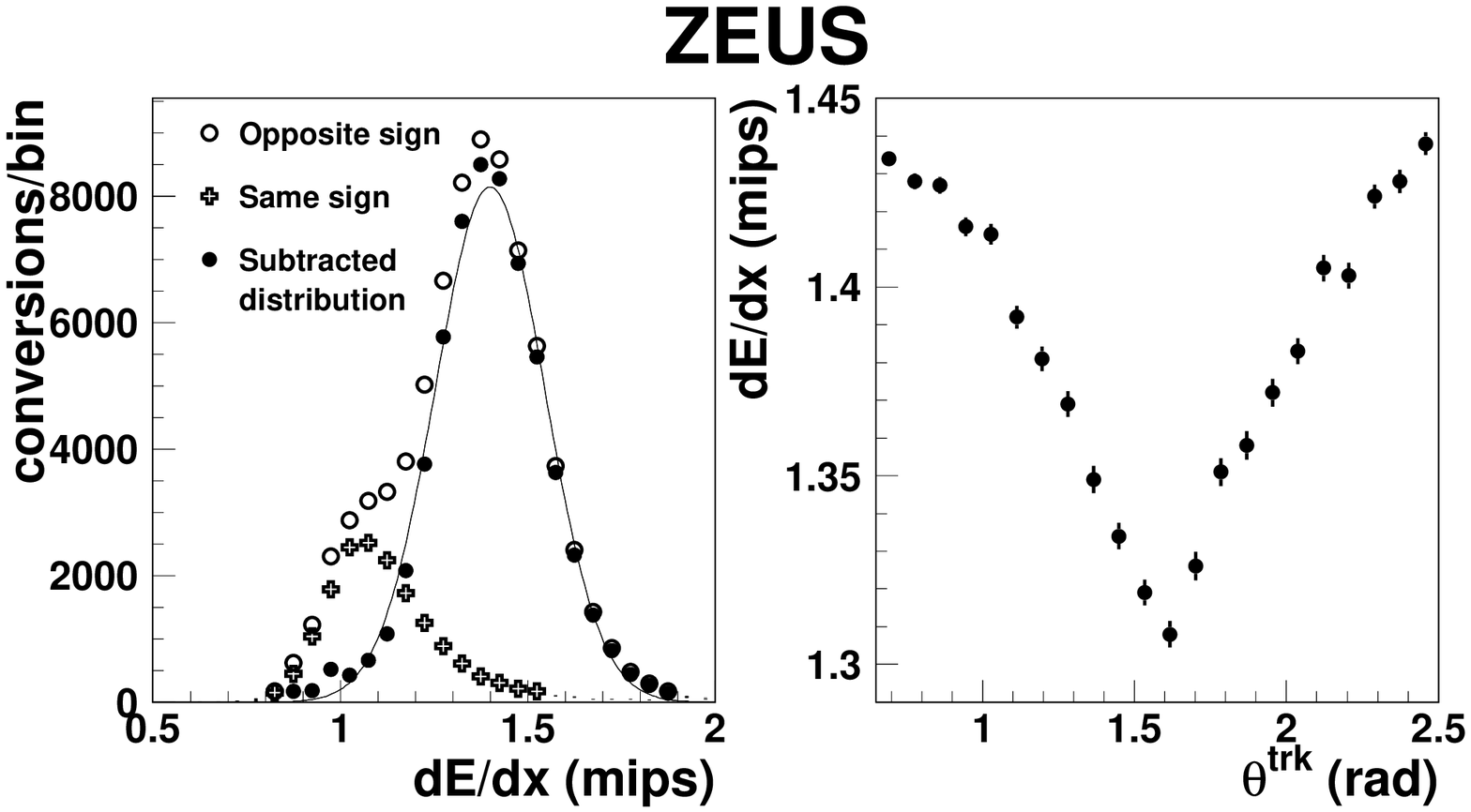,height=8.25cm}
\end{center}
\begin{figure}[htb]
\caption{(a) The $dE/dx$ distribution of photon-conversion candidates and (b) the  
dependence on the polar angle, $\theta^{\rm trk}$, for electrons. In (a), photon-conversion 
candidates having two tracks of opposite charge are shown as open circles whereas those 
with tracks of the same sign are shown as the crosses; the solid circles show the difference 
between these two distributions. A Gaussian fit is shown in (a) for illustration.}
\label{fig:spq}
\end{figure}

\newpage

\centerline{\epsfig{file=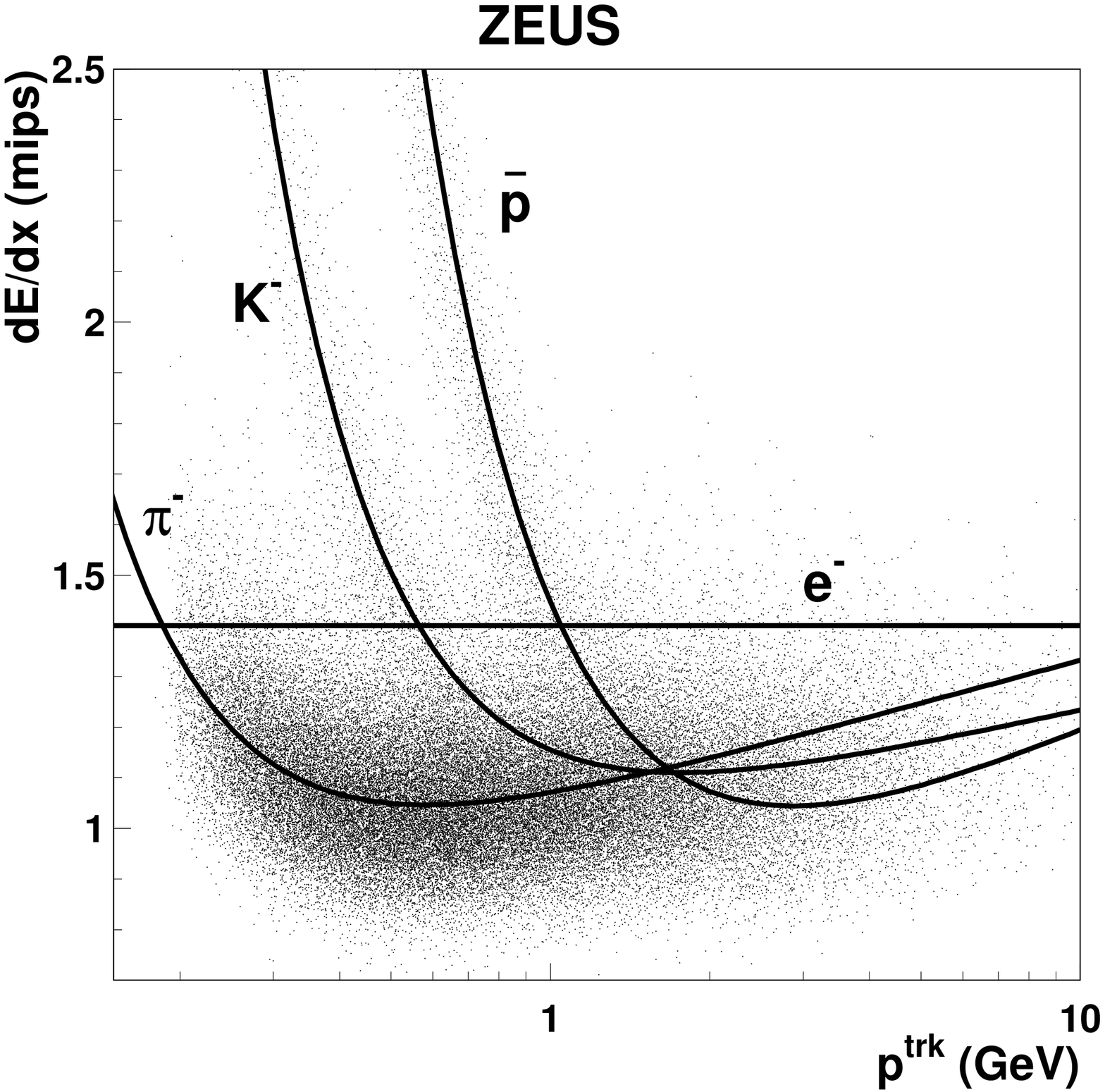,height=17cm}}
\begin{figure}[htb]
\caption{The measured distribution of $dE/dx$ against momentum, $p^{\rm trk}$, for negative 
tracks in the range $|\eta^{\rm trk}|~<~1.1$, as for the analysis. The curves show the 
expected average values for particular types of particles as derived from the Bethe-Bloch 
formula~\cite{bb}. The events are a sub-sample of 
those that pass the dijet trigger requirements.}
\label{fig:dedx-p}
\end{figure}

\newpage

\centerline{\epsfig{file=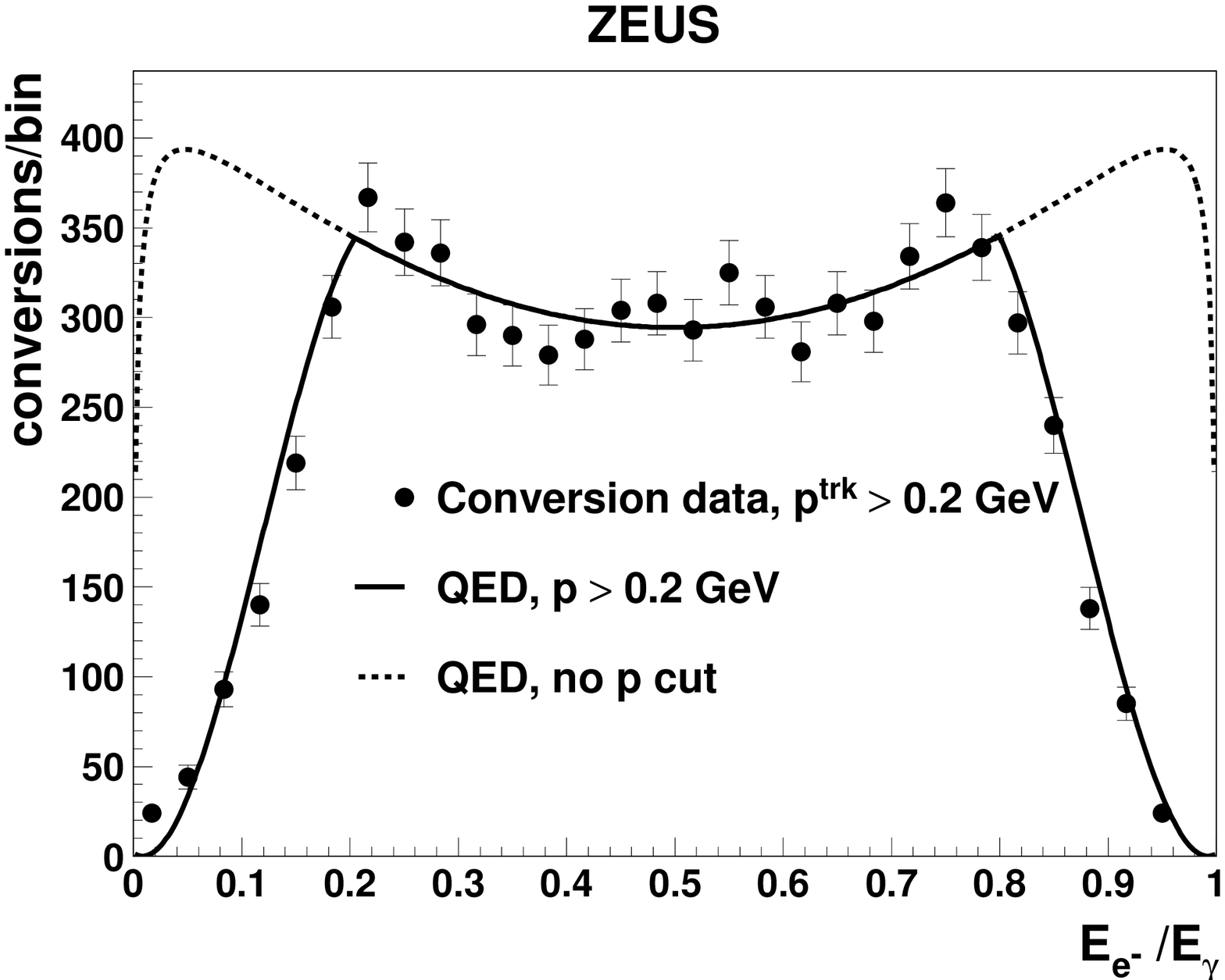,height=12cm}}
\begin{figure}[htb]
\caption{Comparison of the number of electrons found in the data (points) with the 
prediction from pair production (solid line) for conversions in which both tracks have a 
momentum greater than 0.2~GeV. The prediction for no cut on the momentum of the tracks 
is also shown as the dashed line.}
\label{fig:tsai}
\end{figure}

\newpage

\unitlength=1mm
\begin{picture}(0,0)(100,100)
\put(230,60){\bf \Large{(a)}}
\put(230,-19){\bf \Large{(b)}}
\end{picture}
\begin{center}
~\epsfig{file=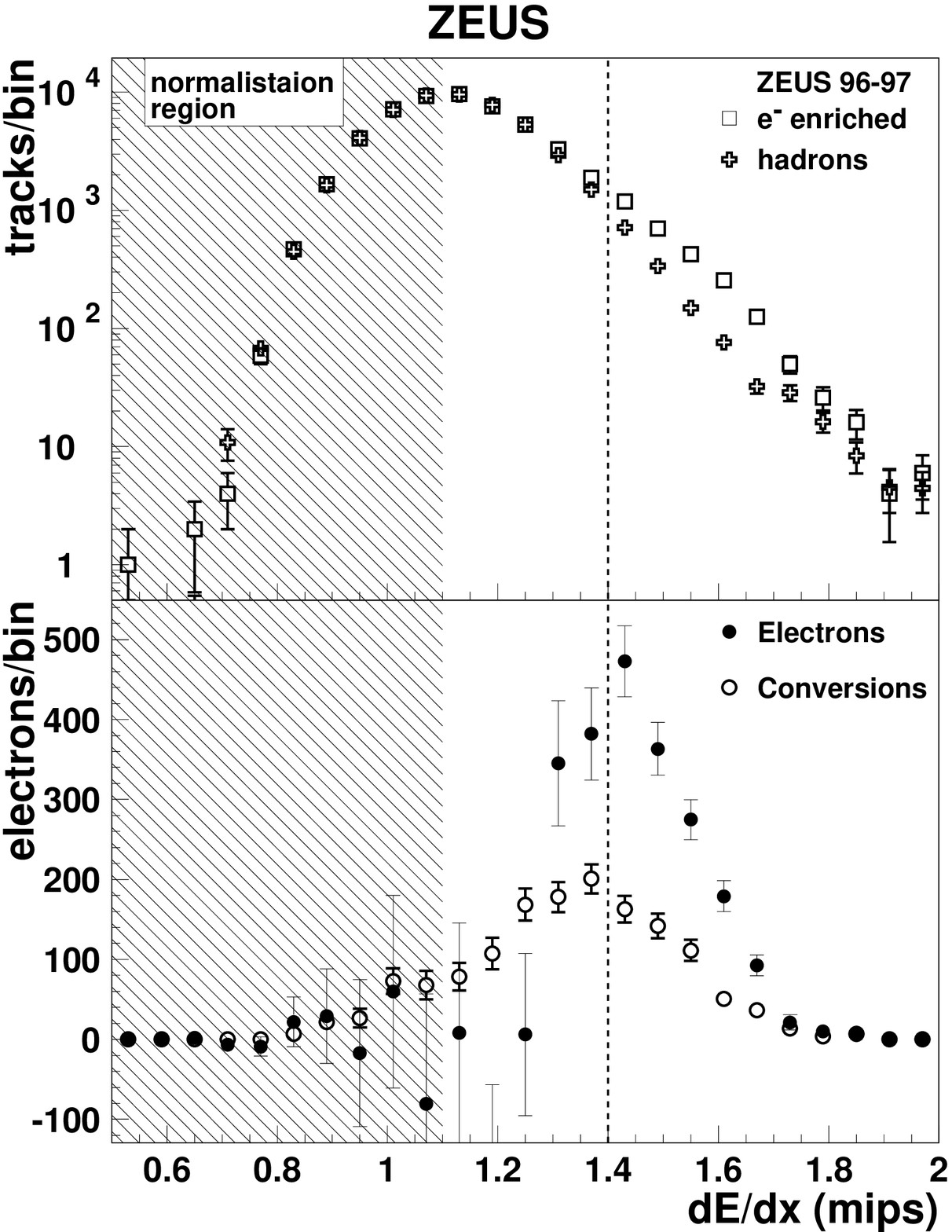,height=17.5cm}
\end{center}
\begin{figure}[htb]
\caption{(a) The $dE/dx$ distribution for the hadronic sample (crosses) and electron-enriched 
(open squares) sample normalised to each other in the hatched region shown. (b) The 
difference between the electron-enriched and hadronic samples (solid circles), together with the 
background arising from photon conversions (open circles). The data in the region with 
$dE/dx>1.4$~mips 
shown by the dashed line were used to extract the results in this paper.}
\label{fig:signal}
\end{figure}

\newpage

\unitlength=1mm
\begin{picture}(0,0)(100,100)
\put(170,84){\bf \Large{(a)}}
\put(252,84){\bf \Large{(b)}}
\put(170,43){\bf \Large{(c)}}
\put(252,43){\bf \Large{(d)}}
\put(170,4){\bf \Large{(e)}}
\put(252,4){\bf \Large{(f)}}
\put(170,-36){\bf \Large{(g)}}
\put(252,-36){\bf \Large{(h)}}
\end{picture}
\begin{center}
~\epsfig{file=data_mc.eps+b_mc,height=16.5cm}
\end{center}
\begin{figure}[htb]
\vspace{-0.5cm}
\caption{Comparison of data (points) with {\sc Herwig} (solid line) and {\sc Pythia} (dashed 
line) MC expectations for: (a) jet transverse energy, $E_T^{\rm jet}$; (b) pseudorapidity of 
the jet, $\eta^{\rm jet}$; (c) number of jets in the event, ${N^{\rm jet}}$; (d) the 
Jacquet-Blondel estimator of $y$, $y_{\rm JB}$; (e) the separation of the electron track and 
jet, $R_{\rm e-jet}$; (f) the distance of closest approach of the track-cluster pair, $d$; 
(g) the transverse momentum of the track, $p_T^{\rm trk}$, and (h) the pseudorapidity of the 
track, $\eta^{\rm trk}$. The MC prediction is area normalised to the number of data events. The 
contribution of events with an electron from the semi-leptonic decay of a $b$ 
quark to the MC prediction is shown as the hatched histogram. The normalisation for the MC from $b$-decays is the 
same as for the total sample and the fraction is that predicted in {\sc Herwig}.}
\label{fig:datamc}
\end{figure}

\newpage

\newpage

\newpage

\end{document}